\documentclass[aps,prl,twocolumn,showpacs,superscriptaddress]{revtex4}

\usepackage{graphicx}
\usepackage{dcolumn}
\usepackage{bm}
\usepackage{color}

\begin{document}

\title{Evidence for a direct band gap in the topological insulator Bi$_2$Se$_3$ from theory and experiment}

\author{I.~A. Nechaev}
 \affiliation{Tomsk State University, 634050, Tomsk, Russia\\}
 \affiliation{Donostia International Physics Center (DIPC), 20018 San Sebasti\'an/Donostia, Basque Country, Spain\\}

\author{R.~C. Hatch}
 \affiliation{Department of Physics and Astronomy, Interdisciplinary Nanoscience Center, Aarhus University, 8000 Aarhus C, Denmark}

\author{M. Bianchi}
 \affiliation{Department of Physics and Astronomy, Interdisciplinary Nanoscience Center, Aarhus University, 8000 Aarhus C, Denmark}

\author{D. Guan}
 \affiliation{Department of Physics and Astronomy, Interdisciplinary Nanoscience Center, Aarhus University, 8000 Aarhus C, Denmark}

\author{C. Friedrich}
 \affiliation{Peter Gr\"{u}nberg Institut and Institute for Advanced Simulation, Forschungszentrum J\"{u}lich and JARA, D-52425 J\"{u}lich, Germany}

\author{I. Aguilera}
\affiliation{Peter Gr\"{u}nberg Institut and Institute for Advanced Simulation, Forschungszentrum J\"{u}lich and JARA, D-52425 J\"{u}lich, Germany}

\author{J.~L.~Mi}
\affiliation{Center for Materials Crystallography, Department of Chemistry, Interdisciplinary Nanoscience Center, Aarhus University, 8000 Aarhus C, Denmark}

\author{B.~B.~Iversen}
\affiliation{Center for Materials Crystallography, Department of Chemistry, Interdisciplinary Nanoscience Center, Aarhus University, 8000 Aarhus C, Denmark}

\author{S. Bl\"{u}gel}
 \affiliation{Peter Gr\"{u}nberg Institut and Institute for Advanced Simulation, Forschungszentrum J\"{u}lich and JARA, D-52425 J\"{u}lich, Germany}

\author{Ph. Hofmann}
 \affiliation{Department of Physics and Astronomy, Interdisciplinary Nanoscience Center, Aarhus University, 8000 Aarhus C, Denmark}

\author{E.~V. Chulkov}
 \affiliation{Donostia International Physics Center (DIPC), 20018 San Sebasti\'an/Donostia, Basque Country, Spain\\}
 \affiliation{Departamento de F\'{\i}sica de Materiales UPV/EHU, Facultad de Ciencias Qu\'{\i}micas, UPV/EHU, Apdo. 1072, 20080 San Sebasti\'an/Donostia, Basque Country, Spain\\}
 \affiliation{Centro de F\'{\i}sica de Materiales CFM - MPC, Centro Mixto CSIC-UPV/EHU, 20080 San Sebasti\'an/Donostia, Basque Country, Spain\\}

\date{\today}

\begin{abstract}
Using angle-resolved photoelectron spectroscopy and \textit{ab-initio} $GW$ calculations, we unambiguously show that the widely investigated three-dimensional topological insulator Bi$_2$Se$_3$ has a direct band gap at the $\Gamma$ point. Experimentally, this is shown by a three-dimensional band mapping in large fractions of the Brillouin zone.  Theoretically, we demonstrate that the valence band maximum is located at the $\Gamma$ point only if many-body effects are included in the calculation. Otherwise, it is found in a high-symmetry mirror plane away from the zone center.
\end{abstract}


\maketitle

Bismuth selenide has been widely studied for many years for its potential applications in optical recording systems \cite{Watanabe1983},
photoelectrochemical \cite{Waters2004} and thermoelectric devices \cite{Mishra1997,Bayaz2003}, and is nowadays commonly used in refrigeration and power generation. Recently, it has attracted increasing interest after its identification as a prototypical topological insulator (TI) \cite{Zhang_NatPhys_2009,Xia_2009_NatMat}. Its surface electronic structure consists of a single Dirac cone around the surface Brillouin zone (SBZ) centre $\bar{\Gamma}$, with the Dirac point (DP) placed closely above the bulk valence band states. In order to exploit the multitude of interesting phenomena associated with the topological surface states \cite{Hsieh_Nature_2009,Hasan_2010_RMP}, it is necessary to access the topological transport regime,
in which the chemical potential is near the DP and simultaneously in the absolute bulk band gap. Due to the close proximity of the DP and the bulk valence states at $\bar{\Gamma}$, this is only possible if there are no other valence states in Bi$_2$Se$_3$ with energies close to or higher than the DP. Therefore, it is crucial to establish if the bulk valence band maximum (VBM) in bismuth selenide is placed at $\Gamma$ (and thus projected out to $\bar{\Gamma}$) or at some other position within the Brillouin zone (BZ). As the bulk conduction band minimum (CBM) is undisputedly located at $\Gamma$ \cite{Greanya_JAP_2002,Chen_Science_2010}, the question about the VBM location is identical to the question about the nature of the fundamental band gap in this TI, direct or indirect.

The nature of the bulk band gap is thus of crucial importance for the possibility of exploiting the topological surface states in transport, but the position of the VBM in band structure calculations remains disputed. In a linearized muffin-tin orbital method (LMTO) calculation within the local density approximation (LDA), the VBM was found at the $\Gamma$ point, implying that Bi$_2$Se$_3$ is a direct-gap semiconductor \cite{Mishra_JPCM_1997}. Contrarily, by employing the full-potential linearized augmented-plane-wave method (FLAPW) within the generalized gradient approximation (GGA), the authors of Ref.~\onlinecite{Greanya_JAP_2002} have found the VBM to be located on the $Z-F$ line of the BZ, which is lying in the mirror plane. Similar results have been obtained in Ref.~\onlinecite{Yazyev_PRB_2012} with the plane-wave pseudopotential method (PWP) within the LDA. Various density functional theory (DFT) calculations of the surface band structure of Bi$_2$Se$_3$ \cite{Zhang_NatPhys_2009, Hsieh_Nature_2009, Eremeev_JETPLett_2010, Kim_PRL_2011} also indicate that the VBM of bulk bismuth selenide is not located at the BZ center. The inclusion of many-body effects within the $GW$ approximation was shown to change this situation somewhat because it decreases the size of the band gap at the BZ center, leading to two essentially degenerate VB maxima, one along the $Z-F$ line and another one at the $\Gamma$ point \cite{Yazyev_PRB_2012}, the former being a mere 0.02~eV below the latter. Apart from that (and a number of other approximations employed in this work to be discussed below), the valence band dispersion has only been calculated along high-symmetry lines of the BZ, precluding a firm conclusion about the position of the VBM in the three-dimensional BZ. Thus, the nature of the band gap has not yet been determined unequivocally.

\begin{figure*}[tbp]
\begin{center}
\includegraphics[angle=0, scale=0.67]{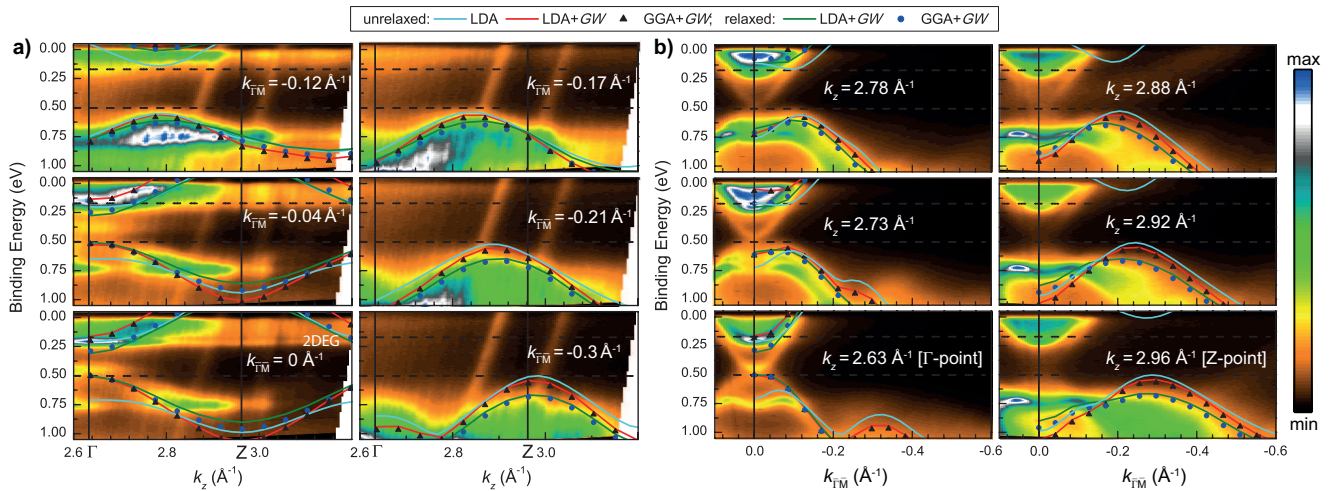}
\caption{(Color online)
Photoemission intensity and theoretical results at different constant parallel momenta $k_{\bar{\Gamma}\bar{M}}$ along directions parallel to $\Gamma-Z$ (a) and constant normal momenta $k_{z}$ along directions parallel to $Z-U$ (b), which are shown by vertical and horizontal dashed lines in Fig.~\ref{fig:2}(a), respectively. Dashed horizontal lines correspond to the experimentally  observed energies of the VBM ($0.505\pm0.030$ eV) and the CBM ($0.170\pm0.025$ eV). The theoretical curves (shifted to have the VBM at the same energy as in the ARPES  experiment) present the lowest conduction band and the uppermost valence band obtained for the experimental \cite{Wyckoff} (unrelaxed) and relaxed atomic positions without (LDA) and with (LDA+$GW$ or GGA+$GW$) the $GW$ corrections to the DFT (LDA or GGA) band structure. The two parallel and nearly vertical lines of higher photoemission intensity seen at $k_{z}$ values of roughly 2.8 and 3.0~$\textrm{\AA}^{-1}$ in (a) correspond to Bi $5d$ core levels which have been excited by  second-order light from the monochromator.}
 \label{fig:1}
\end{center}
\end{figure*}

Experimentally, it appears that the assumption that bismuth selenide is a material with a direct band gap has first been made in Ref.~\onlinecite{Koehler_PSSB_1975}, where galvanomagnetic properties were investigated. By inspecting angle-resolved photoemission spectra in the $\Gamma-Z-F$ direction, the authors of Ref.~\onlinecite{Greanya_JAP_2002} have concluded that the VBM is located at the $\Gamma$ point. More recent experimental studies have focused mostly on the topological state in the immediate vicinity of $\bar{\Gamma}$ (see e.g. Ref. \onlinecite{Xia_2009_NatMat}). A systematic exploration of the experimental dispersion along different lines in the whole mirror plane of the BZ is lacking. Nevertheless, the existing evidence from angle-resolved photoemission spectroscopy (ARPES) points towards a VBM at the $\Gamma$ point, in contrast to most \textit{ab initio} calculations that place the VBM quite far away from $\Gamma$. However, there is an observation that is not consistent with the ARPES results, which appears to be evidence for an indirect band gap coming from scanning tunnelling microscopy (STM) quasiparticle interference experiments \cite{Kim_PRL_2011}.

In this Letter, we report on theoretical and experimental evidence for a direct band gap in bismuth selenide. We systematically explore the valence band structure in large fractions of the BZ and show theoretically that the VBM is found at $\Gamma$, but only if $GW$ corrections are taken into account. The calculated dispersion for the upper VB is found to be in good agreement with ARPES results. Our findings are important for the scattering and transport properties of the topological surface states and may call for a new interpretation of previous results, e.g. the aforementioned quasiparticle interference experiments that were interpreted based on the indirect band gap found in DFT \cite{Kim_PRL_2011}.

Experiments were performed on single crystals of Bi$_{2}$Se$_{3}$ \cite{Bianchi2010Coexistence}. ARPES measurements were carried out on the SGM-3 beamline of the ASTRID synchrotron radiation facility \cite{Hoffmann:2004}. In order to probe the Bi$_{2}$Se$_{3}$ bulk band structure, the photon energy was varied from 14 to 32~eV in 0.1~eV steps.  The combined energy resolution was $\le$~22~meV and the angular resolution was $\sim$0.15$^\circ$. Samples were cleaved \textsl{in-situ} at room temperature, then cooled to $\sim$70~K for measurements. The $k_{z}$ values for the three-dimensional representation of the photon energy scan were calculated using free-electron final states, i.e., $k_{z}=\sqrt{2m_{e}/\hbar^{2}}(V_{0}+E_{\textrm{kin}}\cos^2{(\theta)})^{1/2}$, where $\theta$ is the emission angle and $V_{0}$ is the  inner potential. From the normal-emission ARPES data it is easy to locate a $\Gamma$ point. The value of $V_{0}=11.8$~eV was determined by iteratively changing $V_{0}$ such that the $k_{z}$ value corresponding to this $\Gamma$ point agrees with the size of the Bi$_{2}$Se$_{3}$ BZ.  It  should be noted that $V_{0}=11.8$~eV is in good agreement with values reported previously \cite{Xia_2009_NatMat,Bianchi2010Coexistence,Wray2010Observation,Hatch2011Stability} and is the only value between 1 and 26~eV which results in a  correct $k_{z}$ value for the $\Gamma$ point.

Experimentally, we go beyond the standard approach of ARPES band structure determinations by probing a large fraction of $k$-space on a dense grid of emission angles and photon energies, even though only a fraction of the data in the most relevant  $\bar{\Gamma}-\bar{M}$ direction of the SBZ is shown here. Combined with the assumption of free electron final states, this permits us to plot the photoemission intensity as a function of $k_z$ for a given $k_{\parallel}=k_{\bar{\Gamma}\bar{M}}$ value parallel to the surface along $\bar{\Gamma}-\bar{M}$ [Fig.~\ref{fig:1}(a)] or at constant $k_z$ as a function of $k_{\bar{\Gamma}\bar{M}}$ [Fig.~\ref{fig:1}(b)]. This unconventional way of plotting ARPES results is only possible based on a large data set, but it is excellently suited for a comparison to band structure calculations.

\begin{figure*}[tbp]
\begin{center}
\includegraphics[angle=0, scale=0.95]{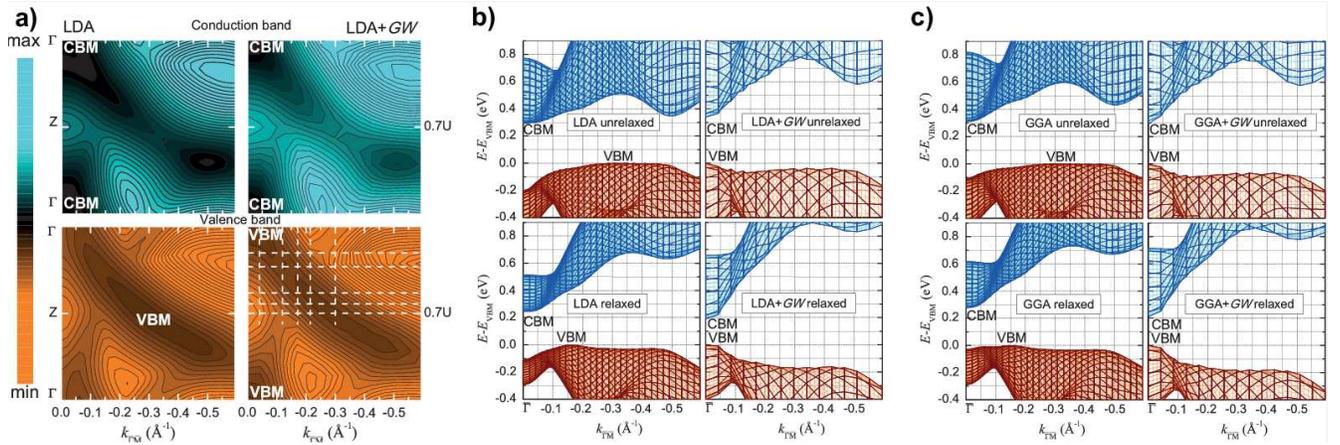}
\caption{(Color online) (a) Contour plots of the lowest conduction band (upper row) and the uppermost valence band (lower row) in the mirror plane (see Fig.~S1 of Supplemental Material \cite{SuppMat}) with the energy spacing of 0.05 eV between contour lines. The presented results are obtained for the experimental atomic positions without (LDA) and with (LDA+$GW$) the $GW$ corrections to the LDA band structure. Vertical and horizontal dashed lines drawn on the valence-band contour plot in the LDA+$GW$ case correspond to the cuts of the three-dimensional photon-energy scan, which are presented in Fig.~\ref{fig:1}. (b) Projections of the lowest conduction band and the uppermost valence band in the mirror plane on the $\bar{\Gamma}-\bar{M}$ direction of the two-dimensional BZ. The presented results are obtained for the experimental (upper row) and relaxed (lower row) atomic positions without (LDA) and with (LDA+$GW$) the $GW$ corrections to the LDA band structure. (c) Same as in (b), but without (GGA) and with (GGA+$GW$) the $GW$ corrections to the GGA band structure.}
 \label{fig:2}
\end{center}
\end{figure*}

Our \textit{ab initio} calculations were performed by employing the FLAPW method as implemented in the FLEUR code\cite{FLEUR} within both the LDA of  Ref.~\onlinecite{LDA_CA_PZ} and the GGA of Ref.~\onlinecite{GGA_PBE} for the exchange-correlation (XC) functional. The two approximations mentioned (LDA and GGA) have been used in order to reveal the effect of different reference one-particle band structures on the $GW$ results. The $GW$ approximation (with the inclusion of spin-orbit interaction as implemented in the SPEX code \cite{SPEX,Sakuma_PRB_2011}) was applied in the one-shot framework, where the Kohn-Sham eigenfunctions are taken as approximate quasiparticle wave functions. As an additional factor that can affect our \textit{ab initio} results, we consider two sets of atomic positions for Bi and Se atoms in a rhombohedral crystal structure. One set of positions corresponds to the experimental atomic positions reported in Ref.~\onlinecite{Wyckoff} (labeled as ``unrelaxed''). Another set (labeled as ``relaxed'') was obtained during a relaxation procedure optimizing the atomic positions at fixed volume. For more computational details, we refer the reader to Sec.~S1 of the Supplemental Material \cite{SuppMat}.

Our $GW$ study goes beyond the one of Ref.~\onlinecite{Yazyev_PRB_2012} in several aspects; instead of the PWP, we employ the FLAPW method, which treats core, valence, and conduction electrons on an equal footing. Furthermore, we do not resort to a plasmon-pole model for the dielectric matrix but directly calculate the full dynamical response within the random-phase approximation (RPA) without any approximation for the frequency dependence. In  Ref.~\onlinecite{Yazyev_PRB_2012} the $GW$ calculation was performed without spin-orbit coupling; spin-orbit interaction was included by second variation using the $GW$ corrected eigenvalues only after the quasiparticle spectrum had been obtained. We use the full four-component spinor wave functions, as obtained from a fully relativistic DFT calculation, directly for the $GW$ calculations as in Ref.~\onlinecite{Sakuma_PRB_2011}. Spin off-diagonal elements in the Green function and the self-energy are thus fully taken into account. Finally, we investigate the behavior of the valence and conduction bands not only on certain lines of the BZ, but in the whole symmetric mirror plane, which is sampled by a dense equidistant mesh composed of 225 {\bf k} points. For each of these points a separate $GW$ calculation was performed, rather than applying Wannier interpolation on a coarse mesh as in Ref.~\onlinecite{Yazyev_PRB_2012}.

The calculations and ARPES results are compared in Fig.~\ref{fig:1}. For the present purpose, the ARPES data should be interpreted as cuts through the spectral function, with the ARPES intensity maximum corresponding to the band structure of the solid. The cut in Fig.~\ref{fig:1}(a), which corresponds to $k_{\bar{\Gamma}\bar{M}} = 0$~$\textrm{\AA}^{-1}$, i.e., normal emission, has two features that disperse along $k_{z}$. These are derived from the CB and VB. The CB is partially occupied by the degenerate bulk doping of the sample. The dispersion of these features is independent of the doping of the sample. Here it is advantageous to use data from a strongly $n$-doped sample such that the VB and CB dispersions are identifiable. However, essentially the same VB dispersion is obtained for an intrinsic sample, or even for a sample with such a strong surface band-bending that both the VB and the CB states are quantized \cite{BianchiSemcon}. This point together with an illustration of how the experimental dispersion is obtained by tracking the intensity maxima in the data are further addressed in Sec.~S2 of the Supplementary Material \cite{SuppMat}.

The experimental data set also shows three features with non-dispersive two-dimensional (2D) character. The first one is the topological state at the Dirac point which is located at a binding energy of $E_{B} \approx 0.45$~eV. The second and third features are the so-called M-shaped state at $E_{B} \approx 0.75$~eV and the parabolic free-electron-like state at $E_{B} \approx 0.25$~eV (the latter is indicated in Fig. \ref{fig:1}(a) as 2DEG -- two-dimensional electron gas). These states have been interpreted as the 2D states formed by the quantization of the valence and conduction band states, respectively, in a potential well formed by the downward bending of the bands at the surface \cite{Bianchi2010Coexistence, King2011Large, Bianchi2011a}. Alternatively, in Refs. \cite{Menshchikova_JETPL_2011, Eremeev_NJP_2012, Ye_ArXiv, Chena_PNAS} the emergence of the states is explained by an expansion of van der Waals spacings due to intercalation of surface-deposited atoms. The intercalation of adsorbed Rb atoms, on the other hand, does not change the surface electronic structure significantly and so this point remains controversial \cite{ASC_NANO_2012}. The photoemission intensity of the surface-related features shows a resonant enhancement at the photon energies for which bulk states with a similar wave function periodicity along $k_z$ is observed. The topological state and the 2D state in the conduction band are enhanced at $\Gamma$ whereas the M-shaped state is enhanced near $Z$.

The best agreement between experimental and theoretical data is reached in the case of the LDA+$GW$ calculations with the relaxed atomic positions. However, the experimental band-gap value ($0.332\pm0.055$ eV) is not so well reproduced as in the LDA+$GW$ case with experimental atomic positions  (see Table \ref{tab:1}). As is clearly seen in Fig.~\ref{fig:1}(a), the shown LDA results, which reflect the situation when the VBM is located in the  mirror plane of the BZ [see Fig.~\ref{fig:2}(a)], are quite far from the experimentally observed valence-band edges. Also, it appears that all calculations overestimate the total width of the upper valence band.

Similar to the conventional semiconductor systems, the $GW$ corrections to the LDA band structure have mainly ``moved'' the conduction band away from
the valence band on the energy scale. This fact is visually represented in Fig.~\ref{fig:2}(a), where one can catch sight of only slightly changed contours
of the bands in the mirror plane upon including the many-body corrections. Nevertheless, there is a crucial point that drastically distinguishes the TI from the conventional semiconductors. This point is the band inversion near the center of the BZ. Owing to this inversion, as was recently shown in Ref.~\onlinecite{Yazyev_PRB_2012}, the mentioned movement apart decreases the ``penetration'' of the bands into each other near the $\Gamma$ point, i.e., the hybridization due to spin-orbit coupling is reduced, and, as a result, both the band-inversion region in $k$-space and the $\Gamma$-point band gap become smaller.

For Bi$_2$Se$_3$, the mentioned mechanism of the many-body corrections finally leads to shifting the VBM to $\Gamma$, as is clearly seen in Fig.~\ref{fig:2}(b), while the VBM of LDA turns into a shallow local maximum that is 0.04~eV below the VBM at $\Gamma$, twice as big a difference as in Ref.~\onlinecite{Yazyev_PRB_2012}. The difference increases further to 0.15~eV when the relaxation of the atomic positions is taken into account as shown in Fig.~\ref{fig:2}(b), which corroborates the identification of the band gap as a direct one. Furthermore, the relaxation causes a larger upward shift of the conduction band beyond the band-inversion region and concomitantly a smaller direct band-gap value for the same reasons as laid out above. On the LDA level, on the other hand, the relaxation results in a shift of the VBM along the mirror plane towards (but not reaching) the center of the BZ and a reduction
of the indirect band gap.

\begin{table}
\caption{\label{tab:1} The calculated band-gap values in eV for Bi$_2$Se$_3$.}
\begin{ruledtabular}
\begin{tabular}{lccc}
        XC   & type     &   unrelaxed        & relaxed \\
  \hline
  LDA          & indirect &     0.30           &  0.25   \\
  LDA+$GW$     & direct   &     0.34           &  0.19   \\
  GGA          & indirect &     0.31           &  0.28   \\
  GGA+$GW$     & direct   &     0.30           &  0.21   \\
\end{tabular}
\end{ruledtabular}
\end{table}

Fig.~\ref{fig:2}(c) shows the GGA band structure without and with the $GW$ corrections. As compared to the LDA-based calculations, the presented data allow one to address the question of how different approximations to the XC functional affect the $GW$ results. We have obtained not only similar behaviour of the bands under study for both reference band structures (LDA and GGA), but  even qualitatively close $GW$-results on the valence-band edge, as one can judge from Fig.~\ref{fig:1}. Besides, the performed DFT+$GW$ calculations with different XC functionals and relaxed atomic positions give practically the same band-gap value (see Table \ref{tab:1}). The difference between the VBM and the local maximum along the $Z-F$ direction increases by 0.03~eV in $GW$ when using a GGA reference system instead of a LDA one. Also, the RPA dielectric constant $\varepsilon_{\infty}$ turns out to be quite stable upon changing
the XC functional and relaxed atomic positions and is consistent with experimental data (see Table SI of the Supplemental Material \cite{SuppMat}).

In conclusion, we have presented ARPES measurements for bismuth selenide which are aimed at a study of the behaviour of the bulk valence-band in the high-symmetry mirror plane in the bulk BZ. We have probed a large range of $k$-space by taking data at a dense mesh of angles and photon energies (only data along $\bar{\Gamma}-\bar{M}$ shown here). We have theoretically considered the effect of many-body corrections within the $GW$ approximation on the band structure,  in particular, on the lowest conduction and highest valence band, and demonstrated how \textit{ab initio} $GW$ results depend on the DFT reference band structure and on atomic positions of Bi and Se atoms in the rhombohedral crystal structure of Bi$_2$Se$_3$. As a net result, we have arrived at theoretical and experimental data which are in good agreement and indicate consistently that Bi$_2$Se$_3$ is a direct-gap semiconductor with the VBM located at the $\Gamma$ point.


\begin{figure*}[tbp]
\begin{center}
\includegraphics[angle=0, scale=0.95]{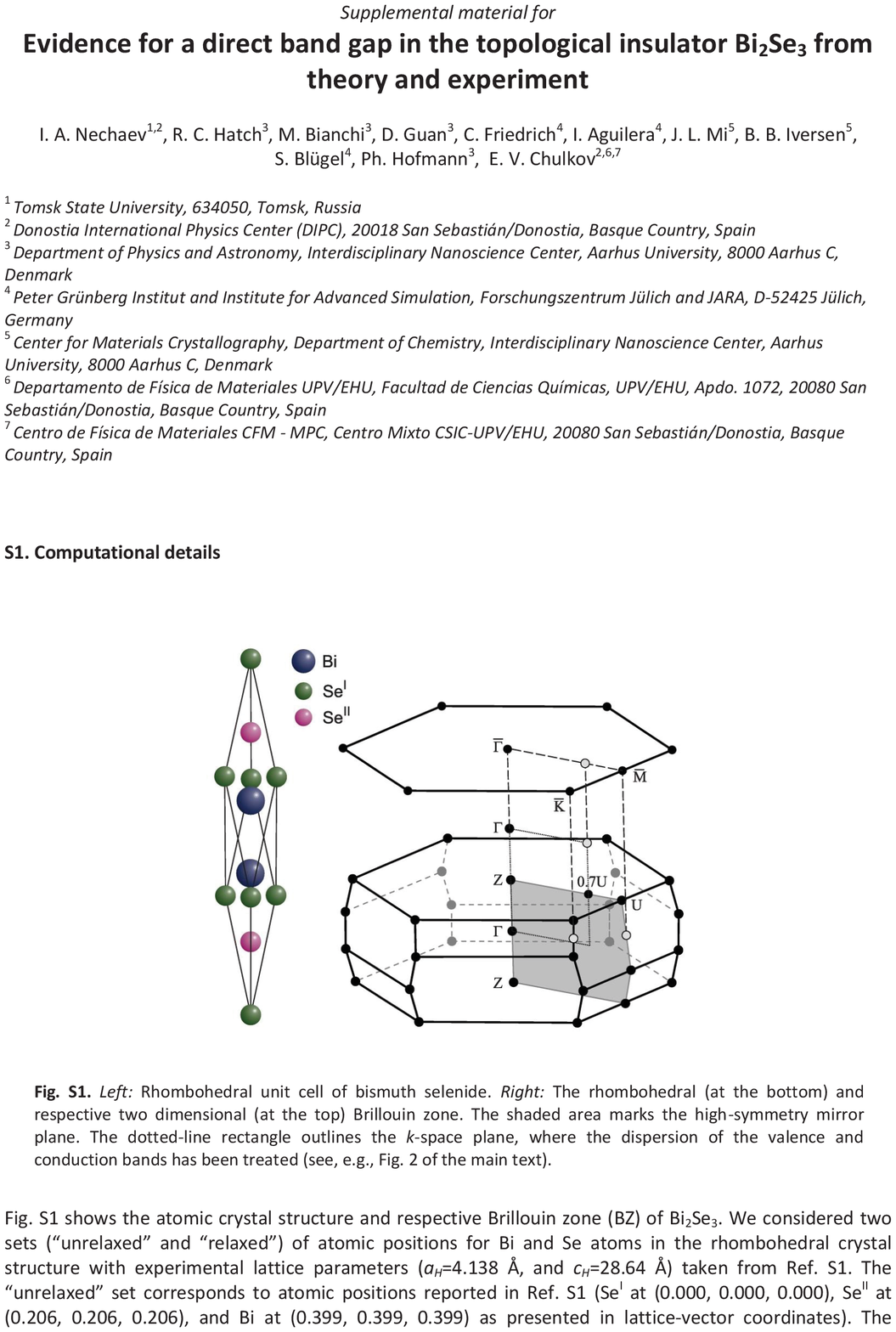}
\end{center}
\end{figure*}
\newpage

\begin{figure*}[tbp]
\begin{center}
\includegraphics[angle=0, scale=0.95]{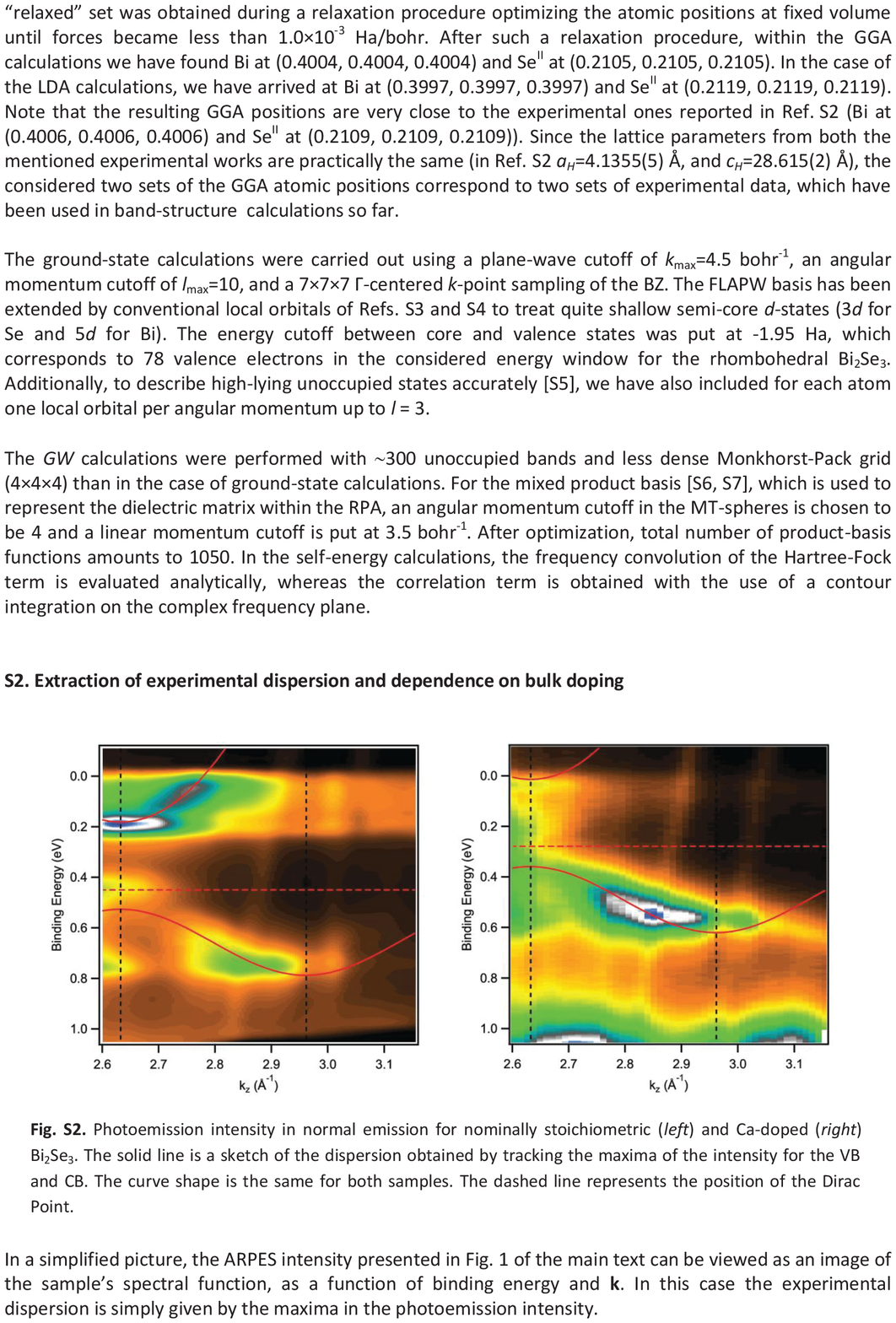}
\end{center}
\end{figure*}

\newpage
\begin{figure*}[tbp]
\begin{center}
\includegraphics[angle=0, scale=0.95]{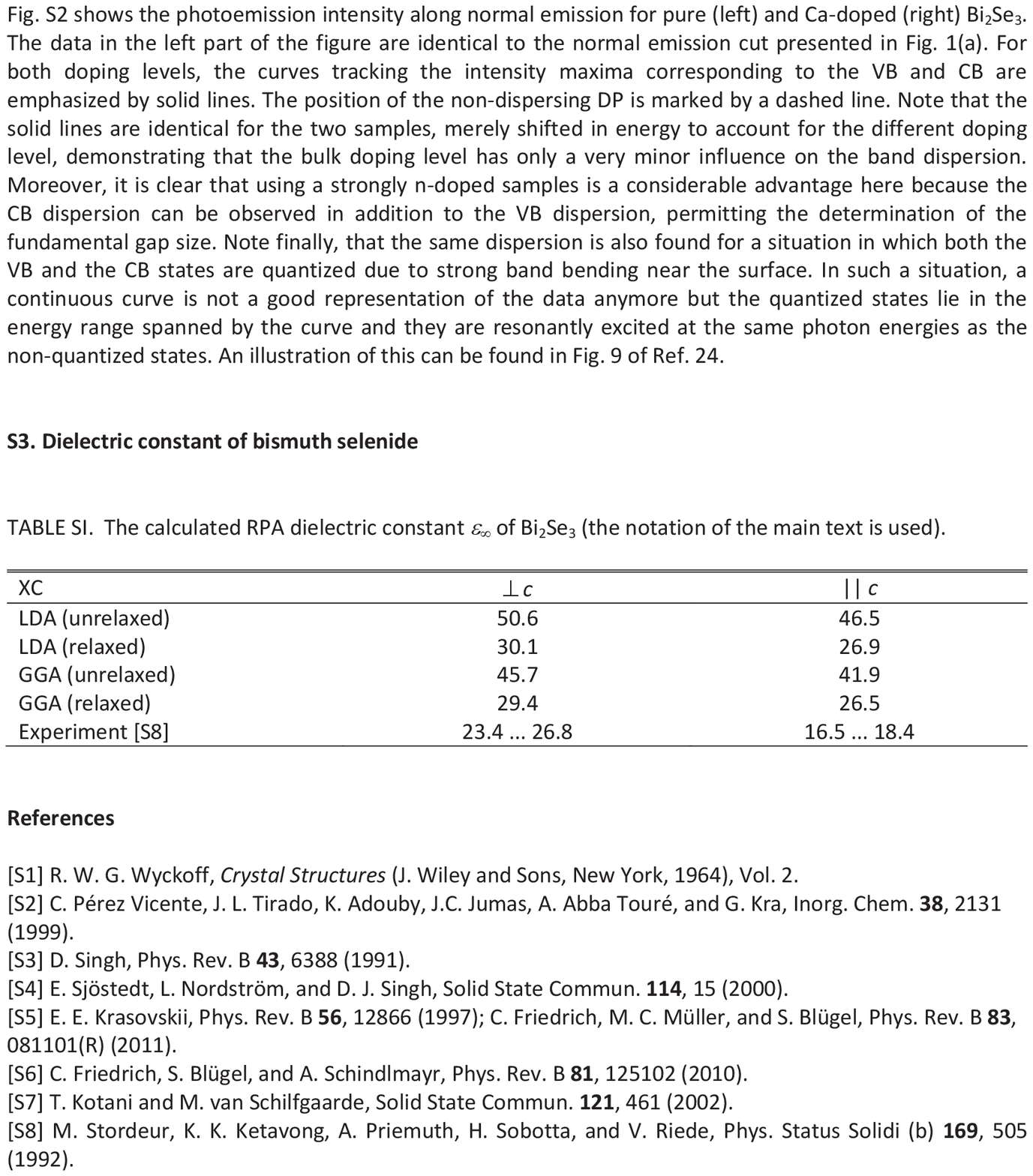}
\end{center}
\end{figure*}


\begin{thebibliography}{99}
%
\bibitem{Watanabe1983} K. Watanabe, N. Sato, S. Miyaoko. J. Appl. Phys. {\bf 54}, 1256 (1983).
%
\bibitem{Waters2004} J. Waters, D. Crouch, J. Raftery, P. O'Brien. Chem. Mater. {\bf 16}, 3289 (2004).
%
\bibitem{Mishra1997} S. K. Mishra, S. Satpathy, O. J. Jepsen. J. Phys. Condens. Matter {\bf 9}, 461 (1997).
%
\bibitem{Bayaz2003} A. A. Bayaz, A. Giani, A. Foucaran, F. Pascal-Delannoy, A. Boyer. Thin Solid Films {\bf 1}, 441 (2003).
%
\bibitem{Zhang_NatPhys_2009} H. Zhang, C.-X. Liu, X.-L. Qi, X. Dai, Z. Fang, and S.-C. Zhang,
Nat. Phys. {\bf5}, 438 (2009).
%
\bibitem{Xia_2009_NatMat} Y. Xia, D. Qian, D. Hsieh, L. Wray, A. Pal, H. Lin, A. Bansil, D. Grauer, Y. S. Hor, R. J. Cava, and M. Z. Hasan, Nature Physics {\bf5}, 398 (2009).
%
\bibitem{Hsieh_Nature_2009} D. Hsieh, Y. Xia, D. Qian, L. Wray, J.H. Dil, F. Meier, J. Osterwalder, L. Patthey, J.G. Checkelsky,
N.P. Ong, A.V. Fedorov, H. Lin, A. Bansil, D. Grauer, Y.S. Hor, R.J. Cava, and M.Z. Hasan, Nature {\bf460},
1101 (2009).
%
\bibitem{Hasan_2010_RMP} M. Z. Hasan C. L. Kane, Rev. Mod. Phys. {\bf82}, 3045 (2010).
%
\bibitem{Greanya_JAP_2002} V. A. Greanya, W. C. Tonjes, R. Liu, C. G. Olson, D.-Y. Chung, and M. G. Kanatzidis, J. Appl. Phys. {\bf92}, 6658 (2002); P. Larson, V. A. Greanya, W. C. Tonjes, R. Liu, S. D. Mahanti, and C. G. Olson, Phys. Rev. B {\bf65}, 085108 (2002).
%
\bibitem{Chen_Science_2010} Y. L. Chen, J.-H. Chu, J. G. Analytis, Z. K. Liu, K. Igarashi, H.-H. Kuo, X. L. Qi, S. K. Mo, R. G. Moore, D. H. Lu, M. Hashimoto, T. Sasagawa, S. C. Zhang, I. R. Fisher, Z. Hussain, and Z. X. Shen, Science {\bf329}, 659 (2010).
%
\bibitem{Mishra_JPCM_1997} S. K. Mishra, S. Satpathy, and O. Jepsen, J. Phys.: Condens. Matter {\bf9}, 461 (1997).
%
\bibitem{Yazyev_PRB_2012} O. V. Yazyev, E. Kioupakis, J. E. Moore, and S. G. Louie, Phys. Rev. B {\bf85}, 161101(R) (2012).
%
\bibitem{Eremeev_JETPLett_2010} S. V. Eremeev, Yu. M. Koroteev, and E. V. Chulkov, JETP Letters, {\bf91}, 387 (2010).
%
\bibitem{Kim_PRL_2011} S. Kim, M. Ye, K. Kuroda, Y. Yamada, E. E. Krasovskii, E. V. Chulkov, K. Miyamoto, M. Nakatake, T. Okuda, Y. Ueda, K. Shimada, H. Namatame, M. Taniguchi, and A. Kimura, Phys. Rev. Lett. {\bf107}, 056803 (2011).
%
\bibitem{Koehler_PSSB_1975} H. K\"{o}hler and A. Fabricius, Phys. Stat. Sol. B {\bf71}, 487 (1975).
%
\bibitem{Bianchi2010Coexistence} M. Bianchi, D. Guan, S. Bao, J. Mi, B. B. Iversen, P. D. C. King, and Ph. Hofmann, Nat. Commun. 1, 128 doi: 10.1038/ncomms1131 (2010).
%
\bibitem{Hoffmann:2004} S. V. Hoffmann, C. S{\o}ndergaard, C. Schultz, Z. Li, and Ph. Hofmann, Nuclear Inst. and Methods in Physics Research, A {\bf523}, 441 (2004).
%
\bibitem{Hatch2011Stability} R. C. Hatch, M. Bianchi, D. Guan, S. Bao, J. Mi, B. B. Iversen, L. Nilsson, L. Hornek\ae{}r, and Ph. Hofmann, Phys. Rev. B {\bf83}, 241303 (2011).
%
\bibitem{Wray2010Observation} L.A. Wray, S.-Y. Xu, Y. Xia, Y.S Hor, D. Qian, A.V. Fedorov, H. Lin, A. Bansil, R.J. Cava, and M.Z. Hasan, Nat. Phys. {\bf6}, 855 (2010).
%
\bibitem{FLEUR} [http://www.flapw.de]
%
\bibitem{LDA_CA_PZ} D.M. Ceperley and B.J. Alder, Phys. Rev. Lett. {\bf45}, 566 (1980) as parametrized by J.P.
Perdew and A. Zunger, Phys. Rev. B {\bf23}, 5048 (1981).
%
\bibitem{GGA_PBE} J.P. Perdew, K. Burke, and M. Ernzerhof, Phys. Rev. Lett. {\bf77}, 3865 (1996); J.P. Perdew, K. Burke, and M. Ernzerhof, Phys. Rev. Lett. {\bf78}, 1396(E) (1997).
%
\bibitem{Sakuma_PRB_2011} R. Sakuma, C. Friedrich, T. Miyake, S. Bl\"ugel, and F. Aryasetiawan, Phys. Rev. B {\bf84}, 085144 (2011).
%
\bibitem{SPEX} C. Friedrich, S. Bl\"{u}gel, and A. Schindlmayr, Phys. Rev. B {\bf81}, 125102 (2010).
%
\bibitem{Wyckoff} R.W.G. Wyckoff, \textit{Crystal Structures} (J. Wiley and Sons, New York, 1964), Vol. 2.
%
\bibitem{SuppMat} See Supplemental Material at http://link.aps.org/supplemental/XXX for computational details, RPA results on the dielectric constant of bismuth selenide, and an extraction of experimental dispersion and dependence on bulk doping.
%
\bibitem{BianchiSemcon} M. Bianchi, R. C. Hatch, D. Guan, T. Planke, J. Mi, B. B. Iversen and Ph. Hofmann, Semicond. Sci. Tech. {\bf27}, 1204001 (2012).
%
\bibitem{King2011Large} P. D. C. King, R. C. Hatch, M. Bianchi, R. Ovsyannikov, C. Lupulescu, G. Landolt, B. Slomski, J. H. Dil, D. Guan, J. L. Mi, E. D. L. Rienks, J. Fink, A. Lindblad, S. Svensson, S. Bao, G. Balakrishnan, B. B. Iversen, J. Osterwalder, W. Eberhardt, F. Baumberger, and Ph. Hofmann, Phys. Rev. Lett. {\bf107}, 096802 (2011).
%
\bibitem{Bianchi2011a} M. Bianchi, R. C. Hatch, J. Mi, B. B. Iversen, and Ph. Hofmann, Phys. Rev. Lett. {\bf107}, 086802 (2011).
%
\bibitem{Eremeev_NJP_2012} S. V. Eremeev, M. G. Vergniory, T. V. Menshchikova, A. A. Shaposhnikov, and E. V. Chulkov, New Journal of Physics {\bf14}, 113030 (2012).
%
\bibitem{Menshchikova_JETPL_2011} T. V. Menshchikova, S. V. Eremeev, and E. V. Chulkov, JETP Letters {\bf94}, 106 (2011).
%
\bibitem{Ye_ArXiv} M. Ye, S. V. Eremeev, K. Kuroda, M. Nakatake, S. Kim, Y. Yamada, E. E. Krasovskii, E. V. Chulkov, M. Arita, H. Miyahara, et al., ArXiv 1112.5869.
%
\bibitem{Chena_PNAS} C. Chena, S. Hea, H. Wenga, W. Zhanga, L. Zhaoa, H. Liua, X. Jiaa, D. Moua, S. Liua, J. Hea, Y. Penga, Y. Fenga, Z. Xiea, G. Liua, X. Donga, J. Zhanga, X. Wangb, Q. Pengb, Z. Wangb, S. Zhangb, F. Yangb, C. Chenb, Z. Xub, X. Daia, Z. Fanga, and X. J. Zhoua, PNAS {\bf109}, 3694 (2012).
%
\bibitem{ASC_NANO_2012} M. Bianchi, R. C. Hatch, Zh. Li, Ph. Hofmann, F. Song, J. Mi, B. B. Iversen, Z. M. Abd El-Fattah, P. L\"{o}ptien, L. Zhou, A. A. Khajetoorians, J. Wiebe, R. Wiesendanger, and J. W. Wells, ACS Nano {\bf6}, 7009 (2012).


\end{thebibliography}
\end{document}